\documentstyle[11pt,epsf]{article}
\newfont{\subsub}{cmr6}

\newcounter{szk}

\begin{document}

\title{A New Approach to Personal Income Distribution}
\author{
\footnote{e-mail address: ishikawa@kanazawa-gu.ac.jp} Atushi Ishikawa$^a$, 
\footnote{e-mail address: tadao@nanao-c.ac.jp} Tadao Suzuki$^b$ and
\footnote{e-mail address: tomoyose@hep.s.kanazawa-u.ac.jp} Masashi Tomoyose$^c$\\ \\ 
$^a$ The Organization of Core Curriculum Studies, Kanazawa Gakuin University\\
Kanazawa 920-1392, Japan\\
$^b$ Department of Management Information, Nanao Junior College\\ 
Nanao 926-8570, Japan\\
$^c$ Institute for Theoretical Physics, Kanazawa University\\ 
Kanazawa 920-1192, Japan}
\date{\today}
\maketitle

%%%%%%%%%%%%%%%%%%%%%%%%%%%%%%%%%%%%%%%%%%%%%%
%%%       ABSTRACT
%%%%%%%%%%%%%%%%%%%%%%%%%%%%%%%%%%%%%%%%%%%%%%
\begin{abstract}
The results of $R^2$ dynamical random surface model
(2-dimensional quantum gravity with a $R^2$ term) 
are applied to explain the personal income distribution.  
A scale invariance exists if there is not the $R^2$ term in the action.  
The $R^2$ term provides a typical scale and breaks the scale invariance explicitly 
in the low and middle income range.  
A new distribution, Weibull distribution, is deduced from the action analytically
in the low income range, and  
a consistent fitting is obtained in the whole income range.  
Also, we show that the lognormal distribution 
in the middle income range can be understood in this framework.  
\end{abstract}

%%%%%%%%%%%%%%%%%%%%%%%%%%%%%%%%%%%%%%%%%%%%%%
%%%       SECTION
%%%       INTRODUCTION
%%%%%%%%%%%%%%%%%%%%%%%%%%%%%%%%%%%%%%%%%%%%%%
\section{Introduction}
The analysis of the personal income distribution is one of 
the important subjects in the econophysics \cite{MS}.
The first step concerning this research is almost established.
There are some empirical knowledge of it.  
It is believed that the distribution follows a power law in the high income range 
and follows a lognormal distribution in the low-middle income range.
The purpose of this paper is to explain this behavior
from a viewpoint of $R^2$ dynamical random surface (DRS) model.
 
The distribution in the high income range is first discussed 
by V. Pareto in 1897 \cite{Pareto}.  
The probability density function of the personal income $x$ is given 
by the following;
\begin{eqnarray}
 p(x)=Cx^{-(1+\alpha)},  
 \label{eq-pareto}
\end{eqnarray}
where $C$ is a normalization factor. 
Also $\alpha$ is the parameter which characterizes
the profile of the income distribution, so-called Pareto index,
and this power law behavior is known as Pareto law.  

On the other hand, 
it is indicated by R. Gibrat in 1931
that the probability density function in the low-middle range follows a lognormal
distribution as follows \cite{Gibrat};  
\begin{eqnarray}
 p(x)=\frac{1}{x\sqrt{2\pi\sigma^2}}\exp\biggl[ 
 -\frac{\log^2\bigl(\frac{x}{x_0}\bigr)}{2\sigma^2}\biggr],
 \label{eq-gibrat}
\end{eqnarray}
where $x_0$ is a geometric mean value of $x$ and $\sigma^2$ is 
a geometric variance.  

In the early '80s, the above two knowledge are rediscovered \cite{Badger}.  
The income distributions in United States 1935-1936 are analyzed 
in these papers.  
Most of the data follows the lognormal distribution, while the higher 
1\% is governed by the power law.  
Recently, this fact is reconfirmed \cite{ASNOTT} by using the detailed 
Japanese data \cite{URL}.  

The personal income distribution has been an interesting object of study.  
What seems to be lacking is a model
from which we can deduce the features in the whole income range.  
For example there are models employed a stochastic evolution equation. 
It is easy to derive a power law, however,
it is hard to lead a lognormal distribution by using these models \cite{LS}.  
Because the basic idea of these models is 
that they contain no typical scale in the fundamental equations.  
We should comment an exception.  
In recent paper \cite{SFA}, the authors employed a concept of 
complex networks \cite{Watts} and a stochastic evolution 
equation with some interactions.
As a consequence, they succeed to construct a power law and lognormal 
distribution in the whole range.  
There is, however, not an analytic consideration for creation of the lognormal 
distribution but only a phenomenological one by using computer simulations.   

Our viewpoint is simple as follows.  
There is no typical scale concerning to money in the high income range.  
The distribution of it, therefore, should obey a power law.  
On the contrary, there is a typical scale in the low-middle income range, 
and it causes the breaking of the power law distribution.  
As a toy model which can realize above two cases in the same framework, 
we employ $R^2$ DRS model which is one of the methods for studying 
2-dimensional quantum gravity with a typical scale.
2-dimensional gravity without a $R^2$ term has no typical scale, 
unlike 4-dimensional gravity. 
Here $R$ means the scalar curvature of a 2-dimensional surface.  
The dimension of $R$ is (length)$^{-2}$, namely the $R^2$ term 
is dimensionfull.  
The $R^2$ term, therefore, provides a typical scale to 2-dimensional gravity. 
It is well known that this model has two ranges.  
A fractal and self-similar structure, $i.e.$ a power law, is seen in the range
where the typical scale does not affect the theory.
On the other hand, the fractal breaks down
in the other range where the scale has the sense. 
This peculiarity is similar to the profile of the personal income distribution. 

In order to confirm our view in the framework of $R^2$ DRS model, 
it is straight to adopt the money itself as the fundamental variable of theory
not the money possessed by individuals.   
We have once forgotten the concept of an individual 
and pay attention only to the connection of the money in a certain unit.
This connection forms a 2-dimensional surface in DRS model,
and later the money of an individual is formed geometrically.  
 
The outline of this paper is as follows. 
Section \ref{sec-minbu} is a brief review of the formalism upon which 
our work is based.  
$R^2$ DRS is explained and MINBU analysis that is one of the methods 
for studying the property of random surface is described.  
The aspects of continuum quantum $R^2$ gravity in 2-dimensions 
and the two asymptotic forms of the partition function are presented 
in section \ref{sec-R2Gin2D}.  
These two forms correspond to the distribution functions in the high 
and the low income ranges, respectively.   
The interpretation and relation between quantities in $R^2$
DRS model and ones in the personal income are discussed 
in section \ref{sec-map}.  
In section \ref{sec-fitting}, the practical fitting is shown.   
The personal income distributions of United States in the year 1935-1936 and 
Japan in the year 1997-2000 are employed for its purpose.  
Summary and discussion will be given in section \ref{sec-summary}.

%%%%%%%%%%%%%%%%%%%%%%%%%%%%%%%%%%%%%%%%%%%%%%%%%%
%%%       SECTION
%%%       R^2  RANDOM  SURFACE  MODEL  AND  MINBU  ANALYSIS
%%%%%%%%%%%%%%%%%%%%%%%%%%%%%%%%%%%%%%%%%%%%%%%%%%
%\section{$R^2$ dynamical random surface model and MINBU Analysis}
\section{$R^2$ DRS Model and MINBU Analysis}
\label{sec-minbu}
In this section, we review DRS and explain MINBU analysis which 
is one of the methods 
for studying DRS \cite{JM}.  
DRS model is known as dynamical triangulation (DT) \cite{KAD} in particle 
physics, and DRS with a $R^2$ term 
has a fractal and a non-fractal ranges \cite{ITY}.    
In DRS, we consider a 2-dimensional surface.  
This surface is constituted 
by attaching the sides of simplexes (for instance equilateral triangles), 
and all possible constitutions are statistically averaged. 

\begin{figure}[htb]
 \begin{minipage}{0.49\textwidth}
  \begin{center}
   \epsfysize=50mm
   \leavevmode
   \epsfbox{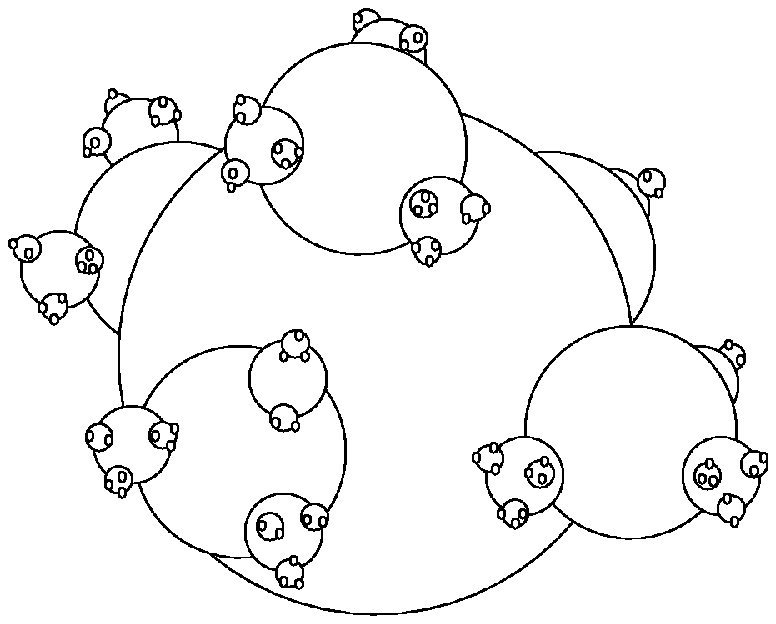}
  \end{center}
  \vspace{-5mm}
  \caption{Fractal 2-dim Random Surface}
  \label{fig:2-dim Random Surface}
 \end{minipage}
 \begin{minipage}{0.49\textwidth}
  \begin{center} 
   \epsfysize=50mm
   \leavevmode
   \epsfbox{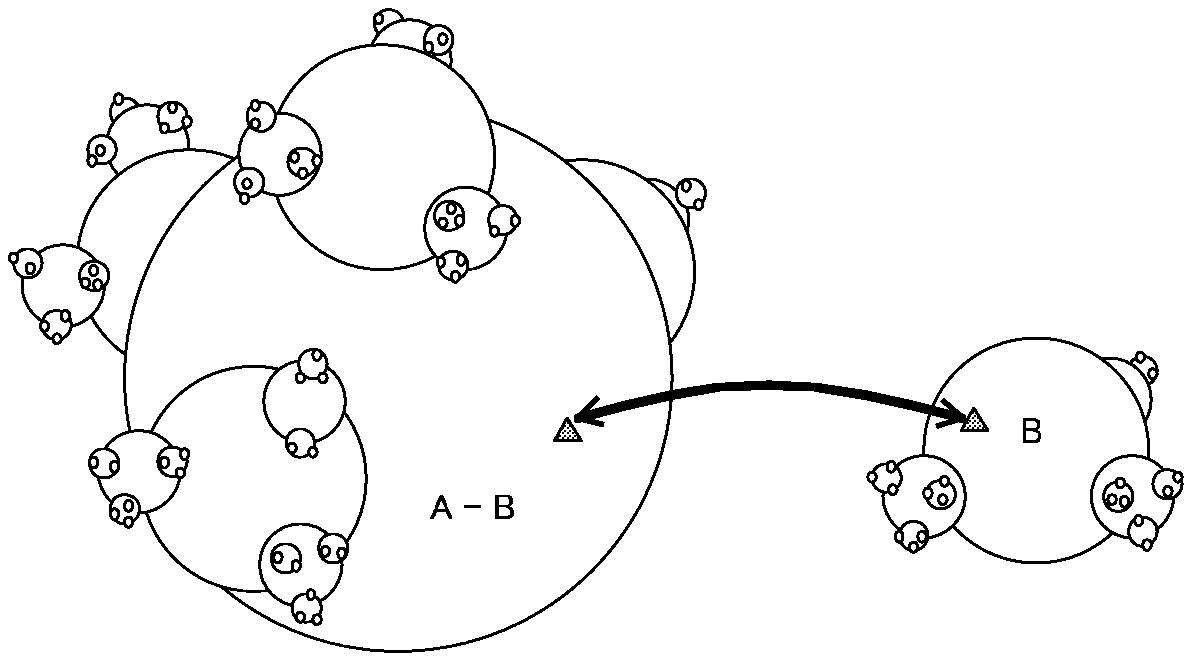}
   \caption{Divided MINBs}
   \label{fig:Divided MINBs}
  \end{center}
 \end{minipage}
\end{figure}

The typical surface is shown in 
Fig.~\ref{fig:2-dim Random Surface}.  
This surface has a self-similar structure.  
A simply connected region of area with one boundary
is called a baby universe.
In the dynamically triangulated surface,
a minimum neck between baby universes is composed of three sides of triangles
and can be identified uniquely.
The distribution of minimum neck baby universe (MINBU) 
is, therefore, one of the important observable quantities in DRS (DT).
In order to analyze it,
we consider the case that a whole universe (area $A$) is divided 
into two baby universes
(area $A-B$, $B$),
paying attention to one of minimum necks (Fig.~\ref{fig:Divided MINBs}).
We denote that $Z(A+1)$ is the fixed area $A+1$ partition function,
and that $Z(A, l)$ is the partition function for surfaces (area $A$) 
with one boundary (length $l$).
MINBU (for minimum boundary loop length $l=3$) can be recognized as a surface 
from which one triangle is removed from a sphere.
We can correlate these two partition functions,
by counting ways which triangle is chosen and removed,
as follows;
\begin{eqnarray}
 Z(A, 3)\sim (A+1)Z(A+1).
\end{eqnarray}
The partition functions of two MINBUs in Fig.~\ref{fig:Divided MINBs} are
given by
\begin{eqnarray}
 Z(B, 3)&\sim& (B+1)Z(B+1),
 \label{B}
\\
 Z(A-B, 3)&\sim& (A-B+1)Z(A-B+1).
 \label{A-B}
\end{eqnarray}
We define, here, that $n_A (B)$ is the statistical average number of MINBUs
of fixed area $B$ on a closed 2-dimensional surface 
of area $A$.
If the 2-dimensional surfaces once divided into two MINBUs 
are made to be one universe again, the following expression
can be obtained; 
\begin{eqnarray}
 n_A(B)\sim \frac{Z(B, 3) Z(A-B, 3)}{Z(A)}.
\end{eqnarray}
This formula works well as long as we treat MINBUs.

It is well known that DT is equivalent to the continuum theory \cite{Takusan}.
If we take the continuum limit 
which brings every sides of triangles to the infinitesimal and
the number of triangles to the infinity simultaneously
keeping the area of a 2-dimensional surface constant, 
it is thought that this limit becomes equivalent to 2-dimensional gravity
defined by the continuous variables.  
The explicit forms of the partition function $Z$ and $n_A(B)$ is obtained 
by using the knowledge 
of 2-dimensional continuum quantum gravity.  
In next section, we consider it.

%%%%%%%%%%%%%%%%%%%%%%%%%%%%%%%%%%%%%%%%%%%%%%%%%%
%%%       SECTION
%%%       CONTINUUM QUANTUM  R^2  GRAVITY  IN  2-DIMENSIONS
%%%%%%%%%%%%%%%%%%%%%%%%%%%%%%%%%%%%%%%%%%%%%%%%%%
\section{Continuum Quantum $R^2$ Gravity in 2-dimensions}
\label{sec-R2Gin2D}
2-dimensional quantum gravity with a $R^2$ term is investigated 
in the continuum 
framework in ref.\cite{KN}.
The partition function for fixed area $A$ is given by
\begin{equation}
Z=\int \frac{{\cal D} g %{\cal D} \chi 
    {\cal D} X}{\rm vol(Diff)}
    {\rm e}^{-S(g)   %_{\chi}(\chi; g)
     -S_M(X^i; g)} 
    \delta \left(\int {\rm d^2}x \sqrt{g}-A \right),
\label{partition}
\end{equation}
where $g_{\mu \nu}(\mu, \nu=0, 1)$ is a metric, $X^i(i=1,2,\cdots ,d)$
is a scalar matter field
and $d$ is the number of scalar fields.
Also $R$ is the scalar curvature of the 2-dimensional surface.  
The $\delta$-function fixes the area to $A$.  
The actions are defined as
\begin{eqnarray}
S(g)&=&\frac{1}{8 \pi} \int {\rm d^2}x \sqrt{g} \left(
    \frac{1}{4 m^2}R^2 + 4 \mu_0 \right),
\label{action chi}
\\
S_M(X^i;g)&=&\frac{1}{8 \pi} \int {\rm d^2}x \sqrt{g} g^{\mu \nu} 
\partial_{\mu} X^i \partial_{\nu} X_i,
\label{action X}
\end{eqnarray}
where $\mu_0$ and $m$ are a bare cosmological constant 
and a coupling constant of length dimension $-1$, respectively.  
The actions (\ref{action chi}), (\ref{action X}) and the integration measure 
${\cal D}g, {\cal D}X$
in the partition function (\ref{partition}) are invariant under 2-dimensional 
diffeomorphisms.
The measure, therefore, should be divided by the volume of the diffeomorphisms,
which is denoted by $\rm vol(Diff)$.
The partition function (\ref{partition}) carried out by an appropriate 
gauge fixing has the exact expression.   
The partition function, however, contains 
infinite excited modes and is difficult to integrate the full functional 
completely.
In Ref.~\cite{KN}, instead of the full integration, the two asymptotic 
forms are shown:
\begin{eqnarray}
Z(A;m,\mu)&\sim& {\rm const.}~A^{\gamma_{\infty }(d)-3} 
{\rm e}^{-\frac{\mu A}{2 \pi}}
                            ~~~~~~~~~~~~~~~~~~
                            {\rm for}~ m^2 A\rightarrow \infty,
\label{infty}                          
\\
&\sim& {\rm const.}~A^{\gamma_0(d, m^2 A)-3} {\rm e}^{-\frac{\mu A}{2 \pi}
                    -\frac{2 \pi}{m^2 A}}
                            ~~~~~~{\rm for}~ m^2 A\rightarrow 0,
\label{0}                               
\end{eqnarray}
where 
$\mu$ is a renormalized cosmological constant, and
\begin{eqnarray}
%\gamma_{\infty }(h, d)&=&\frac{d-25-\sqrt{(25-d)(1-d)}}{12}(1-h)+2,
\gamma_{\infty }(d)&=&\frac{d-25-\sqrt{(25-d)(1-d)}}{12}+2,
\label{gamma_infty}
\\
\gamma_0(d, m^2 A)&=&2+\frac{(d-12)}{6},
\label{gamma_0}
\end{eqnarray}
are string susceptibilities of 2-dimensional gravity.

As for $m^2 A\rightarrow \infty$,
the asymptotic form of the partition function (\ref{infty})
represents that the surface is fractal.
The similar phenomenon can be seen in 2-dimensional gravity 
without the $R^2$ term \cite{BIPZ},
where no typical length scale exists.
In this $A \gg 1/{m^2}$ range, even if the model contains the $R^2$ term,
at an area scale much larger than $1/{m^2}$ surfaces are smoothly fractal.  
On the other hand, as for $m^2 A\rightarrow 0$,
the partition function (\ref{0}) is highly suppressed by an exponential factor
${\rm exp}[-\frac{2 \pi}{m^2 A}]$, hence, the fractal structure
of 2-dimensional surface is broken.
This phenomenon can be understood as follows.
In the $A \ll 1/{m^2}$ range,  
at an area scale much smaller than $1/{m^2}$ surfaces
are affected by the typical length scale, and they are not fractal.

This is the model for which we were asking, however,
the partition function (\ref{infty})
cannot be made to correspond to Pareto law (\ref{eq-pareto}) directly. 
Because area $A$ in Eq.(\ref{infty}) or (\ref{0}) 
should be correspond to the whole money,
on the other hand, $x$ in Eq.(\ref{eq-pareto}) or (\ref{eq-gibrat}) 
is the individual income. 

In order to connect this model to the personal income, we should consider
MINBU analysis in the previous section.
By substituting the asymptotic forms (\ref{infty}), (\ref{0}) to 
the partition function (\ref{B}), (\ref{A-B}),
We can find the following two asymptotic formulae;
\begin{eqnarray}
 n_A(B)&\sim& {\rm const.}~
 %A~
 \biggl[\Bigl(1-\frac{B}{A}\Bigr)B\biggr]^{\gamma _\infty -2}
~~~~~~~~~~~~~~~~~~~~~~~~~~~~~
\mbox{ for $m^2(1-\frac{B}{A})B\rightarrow \infty $},
 \label{eq-n_A-hight}\\
&\sim& {\rm const.}~
\biggl[\Bigl(1-\frac{B}{A}\Bigr)B\biggr]^{\gamma_0-2}
      %\!\!\!
      \exp\biggl[ -\frac{2\pi}{m^2} \frac{1}{\Bigl(1-\frac{B}{A}\Bigr)B}\biggr]
~~   \mbox{ for $m^2(1-\frac{B}{A})B\rightarrow 0 $}.
 \label{eq-n_A-low}
\end{eqnarray}
From these expressions, we can recognize that
the distribution of MINBUs obeys a power law in the limit 
$m^2(1-B/A) B\to\infty$,
on the other hand, it does not follow any power law in the limit
$m^2(1-B/A) B\to 0$.

%%%%%%%%%%%%%%%%%%%%%%%%%%%%%%%%%%%%%%%%%%%%%%%%%
%%%       SECTION
%%%       THE  MAP  BETWEEN  R^2  DRS
%%%       AND  THE  PERSONAL  INCOME
%%%%%%%%%%%%%%%%%%%%%%%%%%%%%%%%%%%%%%%%%%%%%%%%%
\section{The Map between $R^2$ DRS and the Personal Income}
\label{sec-map}
In the previous section, we analyze how the 2-dimensional surface 
(the whole universe)
is distributed to MINBUs.
Let us consider a relation between the quantities of $R^2$ DRS
and those of the personal income.
By comparison between Pareto law~(\ref{eq-pareto}) and 
Eq.~(\ref{eq-n_A-hight}),  
we propose a correspondence relation;
\begin{eqnarray}
 \Bigl(1-\frac{B}{A}\Bigr)B&\Leftrightarrow & x,
 \label{eq-map01}\\
 \frac{n_A(B)}{N}&\Leftrightarrow & p(x),
 \label{eq-map02}
\end{eqnarray}
where $N$ is a normalization factor.
We should note that MINBU is, in general, defined as the smaller universe
when the whole universe is divided into two parts, therefore, $B \le A/2$
($x \le A/4$). From this correspondence, 
we can correlate the string susceptibility to Pareto index
as follows;
\begin{eqnarray}
 \gamma_{\infty}=1-\alpha.
 \label{eq-map_gamma-pareto}
\end{eqnarray}
Since Pareto index $\alpha$ is reported as $1.5$ in \cite{MS, Pareto, Badger}, 
we should take the value of $\gamma_{\infty}$ as $-\frac{1}{2}$.  
This means that we should chose the corresponding 2-dimensional surface 
as a sphere without any matters ($d=0$),
and we can conclude that $\gamma_{0}=0$ by using Eq.~(\ref{gamma_0}).    
Two asymptotic distributions are, in the end, acquired;  
\begin{eqnarray}   
p(x)&\sim& {\rm const.}~x^{-\frac{5}{2}} 
\qquad\qquad\qquad\qquad\qquad\qquad   
\mbox{ for $m^2x\rightarrow \infty $},
     \label{eq-distribution-high}\\
&\sim& 
    {\rm const.}~x^{-2}
  \exp\biggl[-\frac{2\pi}{m^2}\frac{1}{x}\biggr]
\qquad\qquad\qquad~   \mbox{ for $m^2x\rightarrow 0 $}.
     \label{eq-distribution-lowmiddle}
\end{eqnarray}
Here $m$ is a typical area scale.  
We should notice that Eq.~(\ref{eq-distribution-lowmiddle}) is known as 
Weibull distribution.  

The asymptotic distribution (\ref{eq-distribution-lowmiddle}) is correspond to
the low income one, and it is not a
lognormal distribution which is offered by many researchers.  
This fact must be paid attention.  
The form of Eq.~(\ref{eq-distribution-lowmiddle}) obtained by our argument
agrees with the practical data in the low income range.  
Particularly its fitting is better than the lognormal one 
in the low income range.

%%%%%%%%%%%%%%%%%%%%%%%%%%%%%%%%%%%%%%%%%%%%%%%%%%
%%%       SECTION
%%%        THE  DATA FITTING  OF  THE  INCOME  DISTRIBUTIONS
%%%%%%%%%%%%%%%%%%%%%%%%%%%%%%%%%%%%%%%%%%%%%%%%%%
\section{The Data Fitting of the Income Distributions}
\label{sec-fitting}
In this section, we attempt to fit the personal income distributions of Japan 
in several years 1997-2000 \cite{ASNOTT, URL}, 
and the United States in the year 1935-1936 \cite{Badger}.   
For this purpose, we define a cumulative probability 
%to deal with cumulative number 
as follows:
\begin{eqnarray}
 P(x\leq)=\int_x^{\infty}dy\ p(y).
\end{eqnarray}
Substituting Eqs.~(\ref{eq-distribution-high}) and 
(\ref{eq-distribution-lowmiddle}) 
for this definition, 
we obtain below formulae:
\begin{eqnarray}
 P(x\leq)&\sim& C_{\infty}~x^{-\frac{3}{2}} 
  \qquad\qquad\qquad\qquad\qquad~~~~   \mbox{ for $m^2x\rightarrow \infty $},
  \label{eq-distribution-high-cum}
\\
&\sim&  
1- C_0~ \exp\biggl[-\frac{2\pi}{m^2}\frac{1}{x}\biggr]
  \qquad\qquad\qquad  \mbox{ for $m^2x\rightarrow 0 $},
  \label{eq-distribution-lowmiddle-cum}
\end{eqnarray} 
where $C_{\infty}$ and  $C_{0}$ are normalization factors.

\begin{figure}[htb]
 \centerline{\epsfxsize=0.65\textwidth\epsfbox{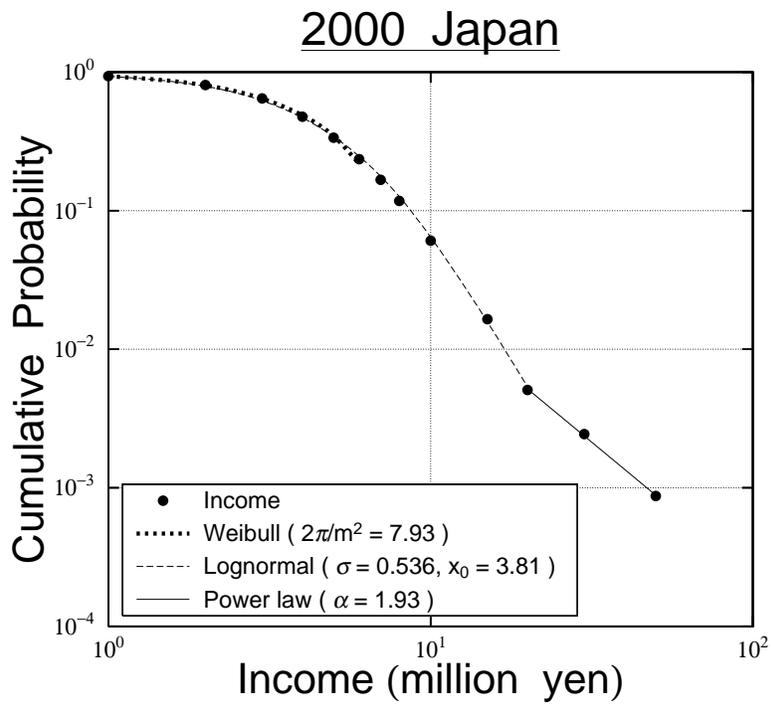}}
 \caption{Cumulative Probability for 2000 Japan}
 \label{fig:2000-japan}
\end{figure}

\begin{figure}[htb]
%%%%%%%%%%%%%%%%%%%%%%%%%%%%%%%%%%%%%%%
%%%%%%%% 1st Line (two pictures)%%%%%%%
%%%%%%%%%%%%%%%%%%%%%%%%%%%%%%%%%%%%%%%
 \begin{minipage}[htb]{0.49\textwidth}
  \epsfxsize = 1.0\textwidth
  \epsfbox{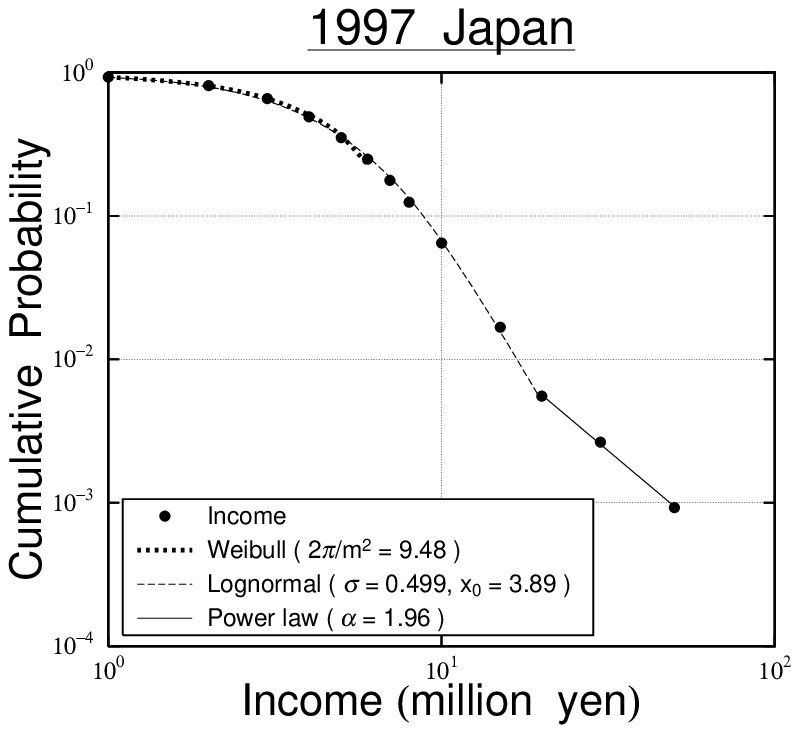}
 \end{minipage}
 \hfill
 \begin{minipage}[htb]{0.49\textwidth}
  \epsfxsize = 1.0\textwidth
  \epsfbox{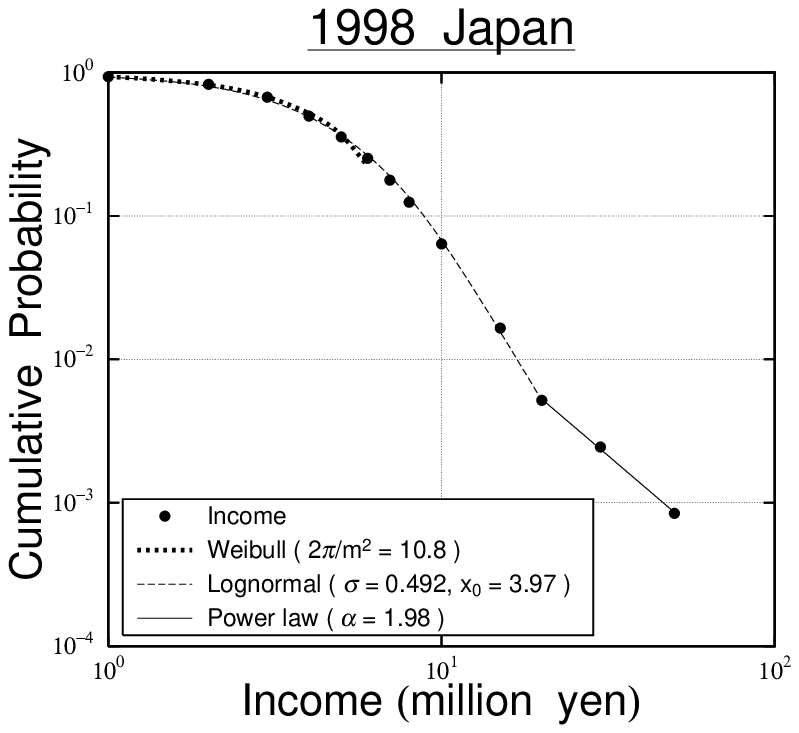}
 \end{minipage}
% \caption{1996 and 1997 Japan Income}
% \label{1996-97-japan}
%\end{figure}
%%%%%%%%%%%% End 1st Line %%%%%%%%%%%%%
%%%%%%%%%%%%%%%%%%%%%%%%%%%%%%%%%%%%%%%
%%%%%%%% 2nd Line (two pictures)%%%%%%%
%%%%%%%%%%%%%%%%%%%%%%%%%%%%%%%%%%%%%%%
%\begin{figure}[htb]
 \begin{minipage}[htb]{0.49\textwidth}
  \epsfxsize = 1.0\textwidth
  \epsfbox{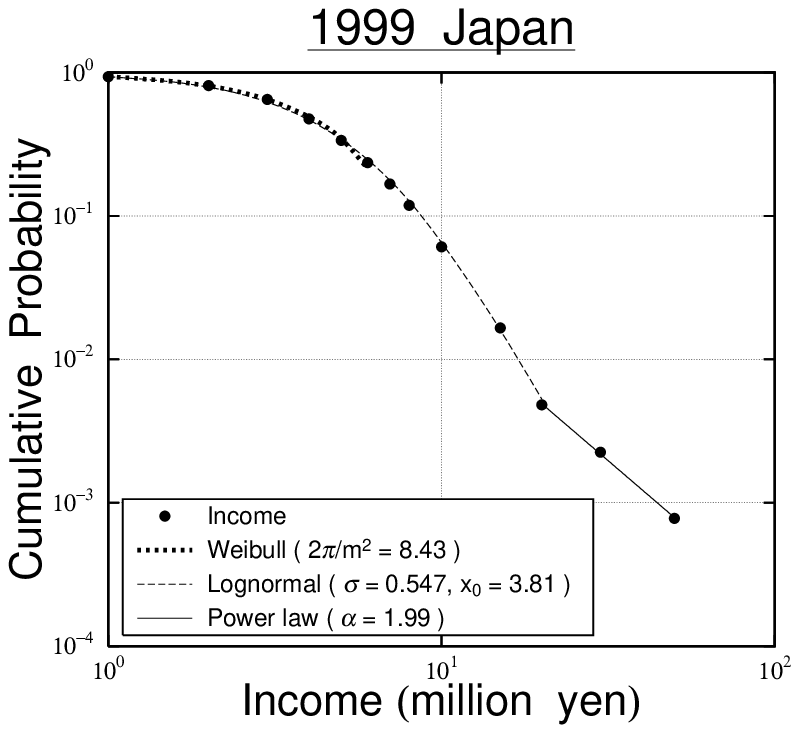}
 \end{minipage}
 \hfill
 \begin{minipage}[htb]{0.49\textwidth}
  \epsfxsize = 1.0\textwidth
  \epsfbox{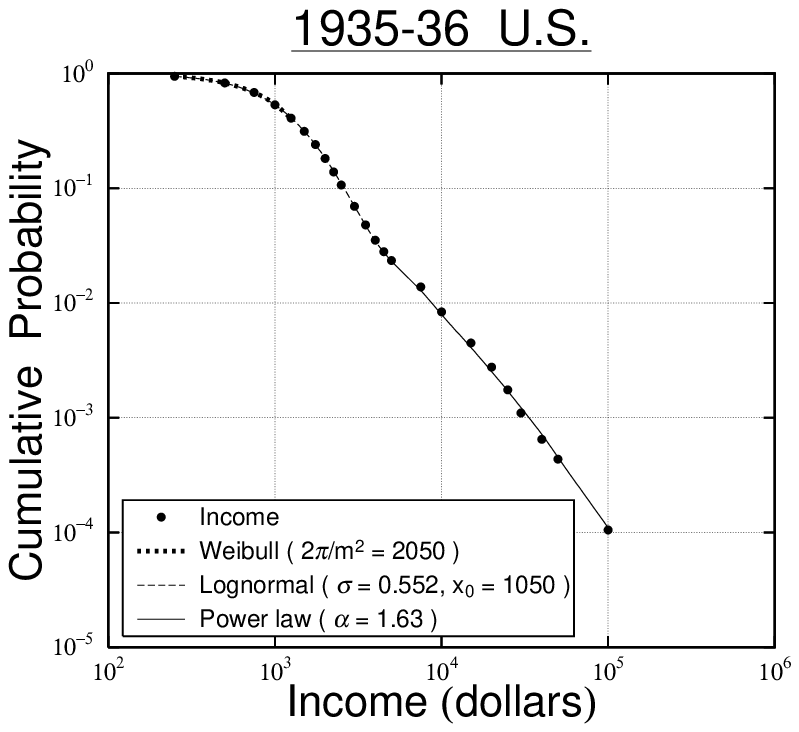}
 \end{minipage}
 \caption{Cumulative Probability for 1997-1999 Japan and 1935-36 U.S.}
 \label{fig:1997-99-japan_1935-36-U.S.}
\end{figure}

Firstly, let us consider the distribution in the high income range.  
In our model, Pareto index $\alpha$ takes the value of $1.5$. 
We find that this index is almost consistent with 
the result of \cite{MS, Pareto, Badger} 
(see U.S. data in Fig.~\ref{fig:1997-99-japan_1935-36-U.S.}).
Also we should notice that this value is little larger than 
that in Japan for the year 1997-2000 \cite{ASNOTT}
(see the Japanese data in 
Figs.~\ref{fig:2000-japan} and~\ref{fig:1997-99-japan_1935-36-U.S.}).  

Secondly, the profile in "the low" income range is considered.  
This paper gives the weight to analyze this distribution.  
It must be noted that "the low" does not mean the low and middle.  
This point of view is one of the main differences from previous works.  
The plots of the data and the best fitting are also presented in 
Figs.~\ref{fig:2000-japan} and~\ref{fig:1997-99-japan_1935-36-U.S.}.  
The parameter $2\pi/m^2$ in the figures can be obtained as the slope 
in the graph 
where the horizontal axis is taken 
as $1/x$ and the vertical axis is taken as 
$\log\bigl(1-P(x\leq)\bigr) $
in the limit $m^2x\to 0$.  

In our model, 
the parameter $m$ specifies the aspect of the breaking of a power law, 
and we are able to make a best fitting by changing the value of it.  
The best fit values of $2\pi/m^2$ in Japan 1997-2000  
and in U.S. 1935-36 are given as 
$9.48$, $10.8$, $8.43$, $7.93$,  
and $2050$, respectively (Figs.~\ref{fig:2000-japan} and 
\ref{fig:1997-99-japan_1935-36-U.S.}).
We have employed the coefficient of determination $r^2$ for the parameter that 
indicates 
the appropriateness of practical fitting:
\begin{eqnarray}
 r^2 = \frac{\Sigma(f_i -\langle F\rangle)^2}
{\Sigma(F_i - \langle F\rangle)^2}.
\end{eqnarray}
Here $F_i$ is the $i$th value of data, $f_i$ is the value of 
the fitting function 
and $\langle F\rangle$ is the mean value of $F_i$.  

\begin{figure}[htb]
 \begin{center}
  \epsfxsize=.8\textwidth
  \leavevmode
  \epsfbox{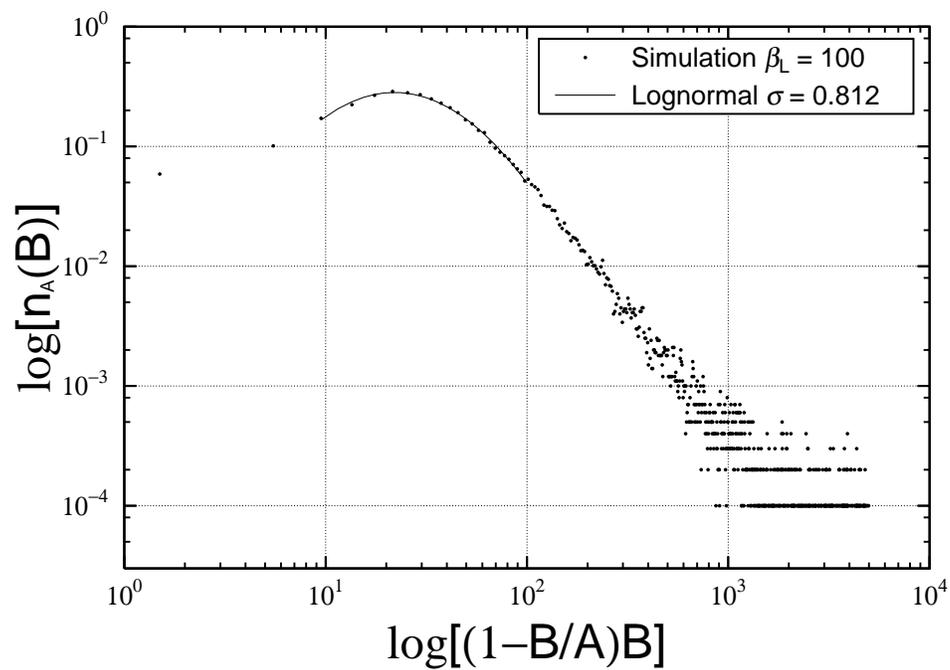}
 \end{center}
 %\vspace{-5mm}
 \caption{Lognormal fitting into the simulation of $R^2$ DRS model}
 \label{fig:Log_simu}
\end{figure}

It remains an unsettled question 
what distribution function should be employed in the middle income range.  
We have to inquire, to some extent, into this subject.  
It is believed that the distribution function is a lognormal one 
in this range.  
It is difficult to deduce a lognormal distribution 
from the partition function (\ref{partition}) analytically.  
We employed, therefore, a computer simulation of DT with a $R^2$ term 
$\beta_L \frac{4\pi^2}{3} \sum_{i} \frac{(6-q_i)^2}{q_i}$ \cite{ITY}
to investigate a profile in the middle income range, 
$i.e. x$ is not so high neither low.
Here $\beta_L$ is a coupling constant and $q_i$ is the number 
of links at the vertex in DRS.  
The result is shown in Fig.~\ref{fig:Log_simu} as the data of the 
distribution of MINBUs
which is not cumulated. 
The horizontal and the vertical axes are
the logarithm of $(1-A/B)B$ and $n_A(B)$, respectively.  
As the figure indicates, 
the profile in the middle range can be well approximated with 
a secondary curve,
and this behavior deduces a lognormal distribution.  
It seems that we can obtain the whole personal income distribution 
from the partition function (\ref{partition}).  

%Fig:: fig:Log_simu

%\newpage
%%%%%%%%%%%%%%%%%%%%%%%%%%%%%%%%%%%%%%%%%%%%%%
%%%       SECTION
%%%       SUMMARY  AND  DISCUSSION
%%%%%%%%%%%%%%%%%%%%%%%%%%%%%%%%%%%%%%%%%%%%%%
\section{Summary and Discussion}
\label{sec-summary}

In this paper we have proposed the $R^2$ dynamical random surface (DRS) 
model as a toy model to explain the profile of the personal income distribution.
The money in a certain unit have been identified with a certain unit area 
(simplex) of a 2-dimensional surface.
The connection of the unit money have been corresponded to the attaching 
of simplexes' sides which forms a 2-dimensional surface.  
On the basis of this correspondence, we have analyzed the distribution of the 
2-dimensional surface statistically as the personal income distribution 
by using MINBU analysis.  

The point is that $R^2$ DRS model has a typical scale $1/m^2$, 
which is the coefficient of the $R^2$ term, 
and that the distribution of MINBUs exhibits two distinct phases.
One is fractal in the large MINBU range where 
the typical scale does not affect the DRS, and
the other is not fractal in the middle-small MINBU range
where small MINBUs are suppressed by the effect of the $R^2$ term.
On the other hand, this fact could be confirmed by the continuum theory
which analytically presented the two asymptotic forms in these two region.  

Moreover, we have proposed that
the personal income $x$ should be corresponded to $(1-B/A)B$,
where $A$ is the whole area of a 2-dimensional surface and $B$ 
is the area of a MINBU. 
From this assumption, the power law distribution has re-obtained in the high 
income range similarly to Refs.~\cite{Pareto, Badger, ASNOTT, LS, SFA}.  
In addition, it has deduced directly from the partition function 
in our method.  
At the same time, we have identified that Pareto index 
$\alpha = 1-\gamma_{\infty}$,
where $\gamma_{\infty}$ is the parameter which characterizes the profile of
the 2-dimensional surface in the limit $x \to \infty$, so called the 
string susceptibility.   
We have also concluded that we should chose the corresponding surface
as a sphere without any matter fields ($d=0$). 
  
We have argued, in particular, the distribution functions in the low and
the middle income range in detail.  
The new distribution function so-called Weibull function has been proposed 
in the low income limit.  
This function has well fitted to the data rather than the lognormal 
distribution.   
A method of least squares has been employed to evaluate a validity 
of the data fitting.
Also we have employed the computer simulation to discussed 
the distribution function  in the middle income range.  
Our whole results were displayed in Fig.~\ref{fig:2000-japan} and 
Fig.~\ref{fig:1997-99-japan_1935-36-U.S.}.  

In our argument, we assume %there was a logical jump 
that the personal income $x$ was identified with the area $(1-B/A)B$. 
We try to explain this identification below.
In MINBU analysis, a boundary of a MINBU is a minimum neck,
where two universes are correlated through only three sides of 
simplexes each other (See Fig.~\ref{fig:Divided MINBs}).
This means that the connection between universes is very weak 
at that point, therefore,
it is natural to consider that the minimum neck separates 
the money which an individual possesses. 

We should notice that the area of a small MINBU is double counted 
as the area of bigger 
one by definition.  
This may be because why the income which an individual obtains
should be added to the income of those who have obtained more income.
The effect $-B/A$ can be neglect when $B$ is much smaller 
than $A$.  
We must, however, take into account of it when $B$ is close to $A/2$.
In our model, there is the huge mother universe
from which all MINBUs are removed (See Fig.~\ref{fig:2-dim Random Surface}).
It is hard to find the object corresponding to this mother universe 
in the real economics, however,
the theoretical distribution obtained has well expressed the real data.
There may be virtual heat bath of the money in the real world. 
These arguments are left behind as a future subject.

In this paper, we have shown that the distribution of the personal 
income could be understandable as
the geometry of a 2-dimensional surface of a sphere without no matter field.
This suggests that the technique developed for the analysis of 
spacetime structure 
can be applied also to the analysis of our social structure.
In 2-dimensions, above geometry is simplest one and
the geometry of a 2-dimensional surface with genus and matter 
fields is also well understood.
There are many fractal objects in the econophysics, for instance 
the distribution of the company profit, 
and the extended geometry may suit these objects.
We are pleased if this analysis becomes a first step for the research 
which considers our society geometrically.

%%%%%%%%%%%%%%%%%%%%%%%%%%%%%%%%%%%%%%%%%%%%%%
%%%%%%%%%%%%%%%%%%%%%%%%%%%%%%%%%%%%%%%%%%%%%%

\section*{Acknowledgements}

The authors would like to express our gratitude to Dr. W. Souma,
Dr. Y. Fujiwara and Professor H. Terao for valuable advices and discussions.  
We are also grateful to Professor H. Kawai for useful comments
especially about his work \cite{KN}. 
Thanks are also due to members of YITP,
where one of the authors (A. I.) stayed several times during the completion 
of this work. 
One of the author (T. S.) is also grateful to Professor Y. Yamagishi 
for useful discussions.

%------------------------------ Bibliography -------------------------

%%%%%%%%%%%%%%%%%%%%%%%%%%%%%%%%%%%%%%%%%%%%
%%%%%%%%%%%%% FIGURE ( PLOTS )%%%%%%%%%%%%%
%%%%%%%%%%%%%%%%%%%%%%%%%%%%%%%%%%%%%%%%%%%%

%\newpage

%%%%%%%%%%%% End 2nd Line %%%%%%%%%%%%%

%%%%%%%%%%%%%%%%%%%%%%%%%%%%%%%%%%%%%%%%%%%%%%
%             FIGURE of Simulation
%%%%%%%%%%%%%%%%%%%%%%%%%%%%%%%%%%%%%%%%%%%%%%

%--------------------------------------------------------------

\end{document}